\documentclass[showpacs,aps,twocolumn,prl]{revtex4}
\usepackage{graphicx}
\begin{document}
\bibliographystyle{apsrev}

\title{Competition between Two Kinds of Correlations in
Literary Texts}

\author{S. S. Melnyk, O. V. Usatenko, V. A. Yampol'skii
\footnote[1]{yam@ire.kharkov.ua}}
\affiliation{A. Ya. Usikov Institute for Radiophysics and Electronics \\
Ukrainian Academy of Science, 12 Proskura Street, 61085 Kharkov,
Ukraine}
\author{V. A. Golick}
\affiliation{Academic Gymnasium No 45, 46-a Tobolskaya Street,
Kharkov, Ukraine}

\begin{abstract}
A theory of additive Markov chains with long-range memory is used
for description of correlation properties of coarse-grained
literary texts. The complex structure of the correlations in texts
is revealed.  Antipersistent correlations at small distances,
$L\alt 300$, and persistent ones at $L\agt 300$ define this
nontrivial structure. For some concrete examples of literary
texts, the memory functions are obtained and their power-law
behavior at long distances is disclosed. This property is shown to
be a cause of self-similarity of texts with respect to the
decimation procedure.
\end{abstract}

\maketitle

\section{Introduction}

The problem of long-range correlated stochastic dynamic systems
(LRSCS) has been under study for a long time in many areas of
contemporary physics~\cite{bul,sok,bun,yan,maj,halvin},
biology~\cite{vossDNA,stan,buld,prov,yul,hao},
economics~\cite{stan,mant,zhang}, etc.~\cite{stan,czir}. An
important examples of complex LRSCS are literary
texts~\cite{schen,kant,kokol,ebeling,uyakm}.

One of the ways to get a correct insight into the nature of
correlations in a symbolic system consists in constructing a
mathematical object (for example, a correlated sequence of
symbols) possessing the same statistical properties as the initial
dynamic system. There exist many algorithms for generating
long-range correlated sequences: the inverse Fourier
transformation~\cite{czir,maks}, the expansion-modification Li
method~\cite{li}, the Voss procedure of consequent random
additions~\cite{voss}, the correlated Levy walks~\cite{shl},
etc.~\cite{czir}. We believe that, of the above-mentioned methods,
the use of the \emph{many-step Markov} chains is one of the most
important because it allows constructing random sequences with
prescribed correlation properties in the most natural way. This
was demonstrated in Ref.~\cite{uya}, where the Markov chains with
the \emph{step-like memory function} (MF) were studied. It was
shown that there exist some dynamical systems (coarse-grained
sequences of the Eukarya's DNA and dictionaries) with correlation
properties that can be properly described by this model.

The many-step Markov chain is the sequence of symbols of some
alphabet constructed using a conditional probability function,
which determines the probability of occurring some definite symbol
of sequence depending on $N$ previous ones. The property of
additivity of Markov chain means the \emph{independent} influence
of different previous symbols on generated one. The concept of
additivity, primarily introduced in paper~\cite{uyakm}, was later
generalized for the case of binary \emph{non-stationary} Markov
chains~\cite{hod}. Another generalization was based on
consideration of Markov sequences with a many-valued
alphabet~\cite{nar1, nar2}.

The efficient method for investigating into the LRSCS systems
consists in decomposing the space of states into a finite number
of parts labelled by definite symbols, which are naturally ordered
according to the dynamics of the system. The most frequently used
decomposition procedure is based on the introduction of two parts
of the phase space. In other words, the approach presupposes
mapping two kinds of states into two symbols, say 0 and 1. This
procedure is often referred to as coarse graining. Thus, the
problem is reduced to investigating the statistical properties of
binary sequences.

It might be thought that the coarse graining could result in
losing, at least, the short-range memory in the sequence. The
authors of Ref.~\cite{nar2} argued that the mapping of a given
sequence into a small-alphabet sequence does not necessarily imply
that the long-range correlations presented in the initial text
would be preserved. However, as was shown in Ref.~\cite{uyakm},
the statistical properties of coarse-grained texts depend, but not
significantly, on the kind of mapping. This implies that only the
small part of all possible kinds of mapping can destroy the
initial correlations in the system. Below, we demonstrate that the
coarse graining retains, although not completely, the correlations
at all distances. This means that there is no point in coding
every symbol (associating every part of the phase space of the
system with its binary code) to analyze the correlation properties
of the dynamic systems, as it is done, for example, in
Ref.~\cite{kokol}, but it is sufficient to use the coarse-graining
procedure.

In the present work, we study the coarse-grained literary texts
examining them as additive Markov chains. A recently obtained
equation~\cite{mel} connecting mutually-complementary
characteristics of these sequence, the memory and correlation
functions, is used. Once the memory function of the original
sequence is found from the analysis of the correlation function,
we construct the corresponding Markov chain with the same
statistical properties. This method for constructing the sequence
of elements with a given correlation function seems to be very
important for other applications, e.g., it can be employed to
fabricate the effective filters of electrical or optical
signals~\cite{iku}.

We show that the memory function of any coarse-grained literary
text is characterized by a complex structure because of the
competition between two kinds of correlations. One type of
correlations works at short distances, $L \alt 300$. The
corresponding MF is negative, which reflects the
\emph{anti-persistent} nature of such correlations. Other type of
correlations with the positive memory function acts at long
distances, $L \agt 300$. The strength of these \emph{persistent}
correlations decreases as a power-law function. We demonstrate
that the power-law decrease of the memory function results in the
self-similarity phenomenon in the coarse-grained texts with
respect to the decimation procedure.

The paper is organized as follows. In the next Section, we
introduce some general relations for the additive Markov chains
and present an equation connecting the correlation and memory
functions. Section III contains the application of the concept of
additive Markov chains to literary works. In Conclusion, we
summarize the obtained results.

\section{Mathematical model}

\subsection{Markov Processes}

Let us consider a homogeneous binary sequence of symbols,
$a_{i}=\{0,1\}$. To determine the $N$-\textit{step Markov chain}
we have to introduce the \emph{conditional probability}
$P(a_{i}\mid a_{i-N},a_{i-N+1},\dots ,a_{i-1})$ of occurring the
definite symbol $a_i$ (for example, $a_i =1$) after $N$-word
$T_{N,i}$, where $T_{N,i}$ stands for the sequence of symbols
$a_{i-N},a_{i-N+1},\dots ,a_{i-1}$. Thus, it is necessary to
define $2^{N}$ values of the $P$-function corresponding to each
possible configuration of the symbols $a_i$ in $N$-word $T_{N,i}$.
Since we will apply our theory to the sequences with long memory
lengthes of the order of $10^6$, some special restrictions to the
class of $P$-functions should be imposed. We consider the MF of
the \textit{additive} form,
\begin{equation}
P(a_{i}=1\mid T_{N,i}) = \sum\limits_{r=1}^{N}f(a_{i-r},r).
\label{1}
\end{equation}
Here the function $f(a_{i-r},r)$ describes the additive
contribution of the symbol $a_{i-r}$ to the conditional
probability of occurring the symbol unity, $a_{i}=1$, at the $i$th
site. The homogeneity of the Markov chain is provided by
independence of the conditional probability (\ref{1}) of the index
$i$. It is possible to consider Eq.~(\ref{1}) as the first term in
expansion of conditional probability in the formal series, where
each term corresponds to the additive (unary), binary, ternary,
and so on functions up to the $N$-ary one.

Let us rewrite Eq.~(\ref{1}) in an equivalent form,
\begin{equation}
P(a_{i}=1\mid T_{N,i})=\bar{a}+\sum\limits_{r=1}^{N}
F(r)(a_{i-r}-\bar{a}), \label{2}
\end{equation}
with
\[
\bar{a}=\frac{\sum\limits_{r=1}^{N}f(0,r)}{[1-\sum\limits_{r=1}^{N}
(f(1,r)-f(0,r))]}
\]
and
\[ F(r)=f(1,r)-f(0,r).\]

We refer to $F(r)$ as the \emph{memory function} (MF). It
describes the strength of influence of previous symbol $a_{i-r}$
$(r=1,...,N)$ upon a generated one, $a_{i}$. It can be shown that
$\bar{a}$ coincides with the value of $a_{i}$ averaged over the
whole sequence. To the best of our knowledge, basically the
concept of the memory function for many-step Markov chains was
originally used in Refs.~\cite{uyakm,uya} where they are well
suited to describe the LRSCS.

The memory function $F(r)$ contains complete information about the
correlation properties of the Markov chain. Usually, the
correlation function and other moments are employed as the input
characteristics describing the correlated random systems. However,
the correlation function describes not only the direct
interconnection of elements $a_i$ and $a_{i+r}$, but also takes
into account their indirect interaction via other intermediate
elements. Our approach operates with the ''origin''
characteristics of the system, specifically with the memory
function. This allows one to disclose the fundamental intrinsic
properties of the system which provide the correlations between
the elements.

A sequence of symbols in a Markov chain can be thought of as the
sequence of states of some particle, which participates in a
correlated Brownian motion. Thus, every $L$-word (a set of
consequent symbols of the length $L$) can be considered as one of
the realizations of the ensemble of correlated Brownian
trajectories in the "time" interval $L$. The positive values of
the MF result in persistent diffusion where previous displacements
of the Brownian particle in some direction provoke its consequent
displacement in the same direction. The negative values of the MF
correspond to the antipersistent diffusion where the changes in
the direction of motion are more probable. Another physical
system, the Ising chain of spins with long-range interactions,
could also be associated with the Markov sequence for which the
positive values of the MF correspond to the attraction of spins
whereas the negative ones conform to the repulsion.

Below we will use some more statistical characteristics of the
random sequences. We consider the distribution $W_{L}(k)$ of the
words of definite length $L$ by the number $k$ of unities in them,
$k_{i}(L)=\sum\limits_{l=1}^{L}a_{i+l}$, and the variance $D(L)$,
\begin{equation}\label{3}
D(L)=\overline{(k-\bar{k})^{2}},
\end{equation}
where the average $\overline{g(k)}$ is defined as
$\overline{g(k)}=\sum\limits_{k=0}^{L}g(k)W_{L}(k)$. Another
important value is the correlation function,
\begin{equation}
K(r)=\overline{a_{i}a_{i+r}}-\bar{a}^{2},\  \
K(0)=\bar{a}(1-\bar{a}). \label{cor}
\end{equation}
By definition, the correlation function is even, $K(r)=K(|r|)$. It
is connected with the above mentioned variance by the
equation~\cite{uyakm},
\begin{equation}
K(r)=\frac{1}{2}(D(r-1)-2D(r)+D(r+1)), \label{3a}
\end{equation}
or
\begin{equation}
K(r)=\frac{1}{2}\frac{d^2 D(r)}{d r^2} \label{3b}
\end{equation}
in the continuous limit.

The memory function used in Refs.~\cite{uyakm,uya} is
characterized by a step-like behavior and is defined by two
parameters only: the memory depth $N$ and the strength $f$ of
symbol's correlations. The value of $f$ was assumed to be
independent of the distance $r$ between the symbols at $r<N$. This
memory function was employed to describe the long-range persistent
properties of the coarse-grained literary texts, specifically, the
super-linear dependence of the variance $D(L)$. However, it does
not reflect the antipersistent behavior of $D(L)$ (observed in
Refs.~\cite{uya}) at short distances. Obviously, we need a more
complex memory function for detailed description of the both
short-range and long-range properties of the coarse-grained texts.

\subsection{Equation for the memory function}

We suggest two methods for finding the memory function $F(r)$ of
the Markov chain $a_i$ that possess the same correlation function
as a given random sequence $b_i$. The first one is based on the
minimization of a ''distance'', $Dist$, between the Markov chain
generated by means of a sought-for MF and the initial sequence
$b_i$. This distance is determined by a formula,
\begin{equation}
\emph{Dist}=\overline{(b_{i}-P(b_{i}=1\mid T_{N,i}))^2}
\label{optim1}
\end{equation}
with $P$-function (\ref{2}). Equating the variational derivative
$\delta \emph{Dist} / \delta F(r)$ to zero, we get the following
relation between the memory function $F(r)$ and the correlation
one $K(r)$:
\begin{equation}
\label{main} K(r)=\sum\limits_{r'=1}^{N}F(r')K(r-r'), \quad r \geq
1.
\end{equation}

Equation~(\ref{main}) can also be obtained by a straightforward
calculation of expression $\overline{a_{i}a_{i+r}}$  in
Eq.~(\ref{cor}) using definition~(\ref{2}) of the memory function.

The second method resulting from the first one establishes a
relationship between the memory function $F(r)$ and the variance
$D(L)$,
\begin{equation}
\label{MF} M(r,0)=\sum\limits_{r'=1}^{N}F(r')M(r,r'), \quad r \geq
1,
\end{equation}
\[
M(r,r')=D(r-r')-(D(-r')+r[D(-r'+1)-D(-r')]).
\]
It is a set of linear equations for $F(r)$ with coefficients
$M(r,r')$ determined by $D(r)$. Equations~(\ref{3a}) and
$D(-r)=D(r)$ are used here.

The function $K(r)$, being a second derivative of $D(r)$, is less
manageable and robust in computer simulations. It is the reason
why we prefer to use the second method (\ref{MF}). This is our
instrument for finding the memory function $F(r)$ of a sequence
using the known variance $D(L)$. The robustness of the method in
the numerical simulations was demonstrated in Ref.~\cite{mel}.

\section{Literature texts viewed as the Markov chains}

\subsection{Variance and correlation function}

\begin{figure}[h!]
\begin{centering}
\scalebox{0.8}[0.8]{\includegraphics{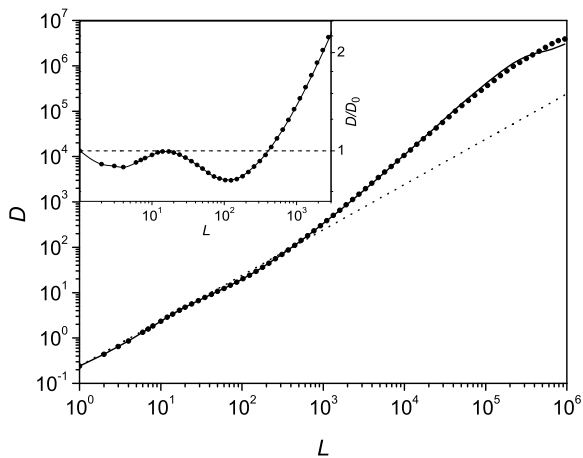}} \caption{The
variance $D(L)$ for the coarse-grained text (letters $(a-m)
\mapsto 0$, letters $(n-z) \mapsto 1$) of the Bible (solid line)
and the Markov chain generated by means of the reconstructed
memory function $F(r)$ (filled circles). The coincidence of these
curves proves the robustness of our method of the MF
reconstruction. The dotted straight line describes the
non-correlated Brownian diffusion, $D_{0}(L)=L\bar{a}(1-\bar{a})$.
The inset demonstrates the antipersistent dependence of the
dimensionless ratio $D(L)/D_0(L)$ upon $L$ at short distances.}
\label{Fig1}
\end{centering}
\end{figure}

\begin{figure}[h!]
\begin{centering}
\scalebox{0.8}[0.8]{\includegraphics{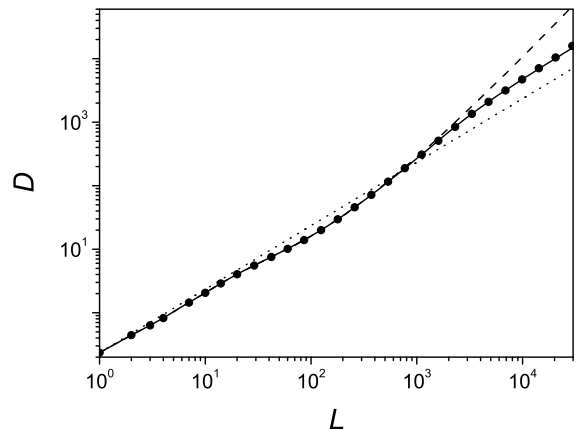}} \caption{The
variance $D(L)$ for the coarse-grained text ((letters with even
numbers in the Alphabet) $\mapsto 1$, (ones with odd numbers)
$\mapsto 0$) of the Bible (dashed line) and for the sequence
obtained by shuffling the blocs of the length $L_{0} = 3000$
(filled circles). The solid line represents the analytical
results, obtained with Eq.~(\ref{corshuf}). The dotted straight
line describes the non-correlated Brownian diffusion,
$D_{0}(L)=L\bar{a}(1-\bar{a})$. } \label{Fig2}
\end{centering}
\end{figure}

Let us apply our method to the investigation into the correlation
properties of the coarse-grained literary texts. At the outset, we
examine the variance $D(L)$ of the coarse-grained text of the King
James Version of the Bible \cite{bibe}. The result of the
numerical simulation is presented by solid line in
Fig.~\ref{Fig1}. The straight dotted line describes the variance
$D_{0}(L)=L\bar{a}(1-\bar{a})$, which corresponds to the
non-correlated biased Brownian diffusion. One of the typical
coarse-graining procedure was used for mapping the letters of the
text onto the symbols zero and unity, ($(a-m) \mapsto 0, (n-z)
\mapsto 1$). It is clearly seen that the diffusion is
antipersistent at small distances, $L\alt 300$, (see inset)
whereas it is persistent at long distances~\cite{border}. The
deviation of the solid line from the dotted one testifies to the
existence of the correlations in the text of the Bible.
\protect
To confirm this statement we break down the original text into
subsequences of a given length $L_{0}=3000$ and randomly shuffle
them. The results from the calculation of the variance for the
coarse-grained initial and shuffled texts of the Bible are given
in Fig.~\ref{Fig2}. For $L\ll L_{0}$, the difference in $D(L)$ is
negligible~\cite{dif}. At $L \sim L_{0}$, the variance and
correlation function of the shuffled sequence are less than the
original ones. At $L>L_{0}$, the correlations in the shuffled text
vanish. In this region, the variance $D(L)$ is a linear function,
and the correlation function being the second derivative of
variance equals to zero.

It is ease to show that the correlation function of the shuffled
sequence can be written as,
\begin{equation}
\label{corshuf} K(r)=\cases {K_{0}(r)(1-\frac{r}{L_{0}}),\qquad
 r < L_{0}, \cr 0,\qquad \qquad \;\;\;\;\;\;\;\;\;\;\;\;\
 r \geq L_{0},}
\end{equation}
where $K_{0}(r)$ is the correlation function of the original
non-shuffled sequence. The corresponding variance obtained by
double numeric integration (see Eq.~(\ref{3b})) of the function
$K(r)$ given by Eq.~(\ref{corshuf}) is shown in Fig.~\ref{Fig2} by
solid line.
\protect
\protect\begin{figure}[h!]
\begin{centering}
\scalebox{0.8}[0.8]{\includegraphics{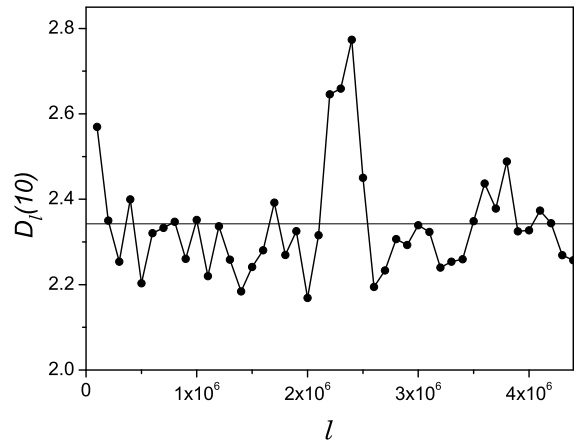}} \caption{The local
variance $D_{l}(10)$ for the coarse-grained text of the Bible vs
the distance $l$. The averaging interval is $L_{0}=10^{5}$.}
\label{Fig3}
\end{centering}
\end{figure}

Along with the global characteristic $D(L)$, it is interesting to
study its local analogue,
\begin{equation}\label{13}
D_{l}(L)=<(k-<k>_{L_{0}})^{2}>_{L_{0}},
\end{equation}
where $L_{0}$ is the interval of local averaging and $l$ is the
coordinate of the left border of this interval. An existence of a
trend in the dependence $D_{l}(L)$ on $l$ would be clearly
indicative of non-stationarity of the stochastic process being
studied. To verify the stationarity of the coarse-grained text of
the Bible we perform the numerical simulation of the $D_{l}(L)$
dependence on $l$ at different fixed values of $L$. As an example,
the result of this simulation for $L=10$ is shown in
Fig.~\ref{Fig3}. It is clearly seen that there exist regular
fluctuations without a pronounced trend. The fluctuations result
from the finiteness of interval $L_{0}$ of averaging. This fact
allows us to make a conclusion about stationarity of the
coarse-grained text of the Bible. The similar analysis of many
other texts gave the same result. It is expedient to study the
global characteristics $D(L)$, Eq.~(\ref{3}), of the sequence
instead of the local one, $D_{l}(L)$.

\subsection{Memory function}

\begin{figure}[h!]
\begin{centering}
\scalebox{0.8}[0.8]{\includegraphics{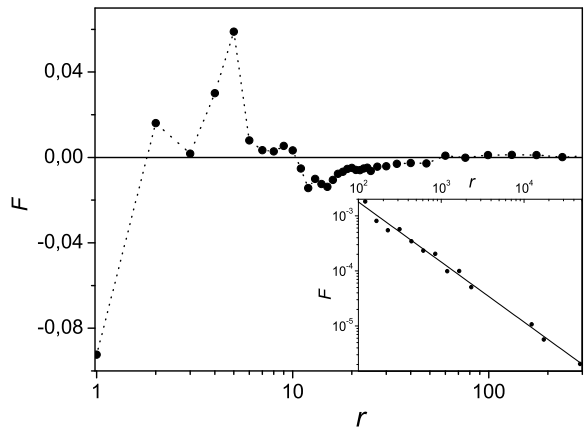}} \caption{The
memory function $F(r)$ for the coarse-grained text of the Bible at
short distances. The power-law decreasing portion of the $F(r)$
plot for the Bible is presented by filled circles in the inset.
The solid line corresponds to the power-law fitting.} \label{Fig4}
\end{centering}
\end{figure}

According to Eqs.~(\ref{main}) and (\ref{MF}), the memory function
can be restored using the variance or the correlation function.
The MF thus obtained for the coarse-grained text of the Bible at
$r<300$ is given in Fig.~\ref{Fig4}. At long distances, $r>300$,
the memory function can be nicely approximated by the power
function $F(r)=0.27r^{-1.1}$, which is shown by the solid line in
the inset in Fig.~\ref{Fig4}. Note that the persistent part of the
MF, $F(r>300)\leq 0.0008$, is much less than its typical magnitude
$0.02$ in the antipersistent region $r<40$.
\protect

It should be emphasized that the short-range part of the memory
function at $r \alt 40$, as well as the $D(L)$ function at $L \alt
300$, is essentially dependent on the method of coarse-graining.
Nevertheless, the antipersistent correlations exist for
practically all kinds of the coarse-graining procedure. An
interesting feature is that the region $r \alt 40$ of negative
antipersistent memory function provides much longer distances $L
\sim 300$ of antipersistent behavior of the variance $D(L)$.

\begin{figure}[h!]
\begin{centering}
\scalebox{0.8}[0.8]{\includegraphics{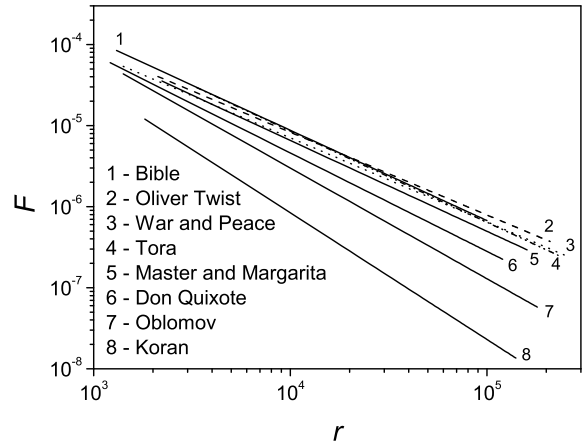}} \caption{The
memory function at long distances for the coarse-grained texts of
eight literary works: 1.~The Bible, 2.~"Oliver Twist" by Charles
Dickens, 3.~"War and Peace" by Leo Tolstoy, 4.~The Tora,
5.~"Master and Margarita" by Mikhail Bulgakov, 6.~"Don Quixote" by
Miguel de Servantes, 7.~"Oblomov" by Ivan Goncharov, 8.~The Koran.
} \label{Fig5}
\end{centering}
\end{figure}

In order to prove the universal character of the power-law
decrease of the memory function at long distances, we compare the
MF of the coarse-grained texts for more than fifty different
literary works. The texts are coarse-grained by mapping the
letters from the first and second halves of the alphabet into zero
and unity, respectively. Subsequently, using Eq.~\ref{MF}, we
first calculate the variances and then the memory functions. All
curves for the memory functions can be well fitted by the
power-law functions $F(r)=cr^{-b}$. The results of the fitting for
eight texts written or translated into Russian~\cite{bibr,libru}
are shown in Fig.~\ref{Fig5}. The exponents in all curves vary
over the interval between $b_{min}=1.02$ for ''War and Peace'' and
$b_{max}=1.56$ for the Koran. Thus, the constants $c$ and $b$ can
be used for linguistic classification of different literary works.
It is interesting to see that the memory functions for the texts
of the English- and Russian-worded Bible, as well as the texts of
the Old and New Testaments are practically coincident.
\protect

The existence of two characteristic regions having different
behavior of the memory function and, correspondingly, of the
persistent and antipersistent portions in the $D(L)$ dependence
appears to be a prominent feature of all texts in any language.
Note that the antipersistent portion of the memory function
corresponds to the region where the grammatical rules are in use.
Therefore, we call this kind of correlations the ''grammatical''
ones. The persistent correlations in a text at very long distances
can be related to a general idea of the literary work. Thus, this
kind of correlations is referred to as the ''semantic'' ones.

\begin{figure}[h!]
\begin{centering}
\scalebox{0.8}[0.8]{\includegraphics{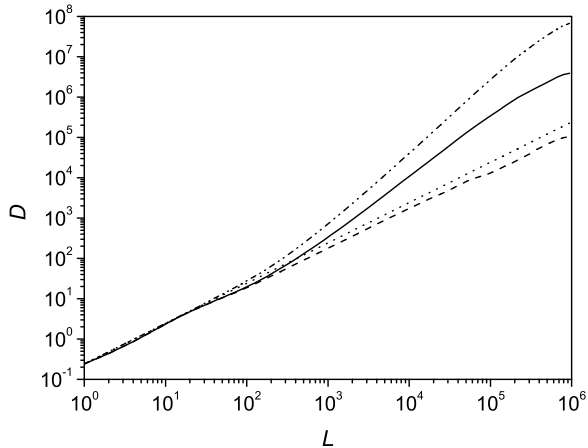}} \caption{The
variance $D(L)$ for the coarse-grained text of the
Bible~\cite{bibe} (the solid line), and for the sequences
constructed with using the persistent part of the MF
(dash-dot-dotted line) and the antipersistent one (dashed line).
The dotted line describes the non-correlated Brownian diffusion,
$D_{0}(L)=L\bar{a}(1-\bar{a})$.} \label{Fig6}
\end{centering}
\end{figure}

Two fundamentally different portions in the MF plots result from a
peculiar competition between the two above-mentioned kinds of
correlations. We would like to stress that both portions of the MF
are equally important to gain an insight into the correlation
properties of the literary texts. To support this statement we
generate two special sequences. In both of them, only one kind of
the memory function for the coarse-grained text of the Bible is
taken into account, and the memory function in another region is
assumed to be zero. The variance $D(L)$ for these two sequences is
given in Fig.~\ref{Fig6}. The lower (dashed) line corresponds to
the case where only the negative antipersistent portion, $r<40$,
of the memory function is allowed for. The upper (dash-dot-dotted)
curve corresponds to the sequence, which is generated by means of
the long-range persistent memory, $F(r)=0.27r^{-1.1}$, $r>100$. It
is evident that the generated sequence with the antipersistent
memory function displays the sub-diffusion only, whereas the
sequence that corresponds to the persistent memory function is
characterized by the super-diffusion behavior of the variance
$D(L)$. The difference between the variances for two generated
sequences and for the original coarse-grained text of the Bible,
shown by the solid line in the same figure, corroborates our
assumption about the significance of both kinds of the memory
function. \protect

\subsection{Self-similarity of the coarse-grained texts}

\begin{figure}[h!]
\begin{centering}
\scalebox{0.8}[0.8]{\includegraphics{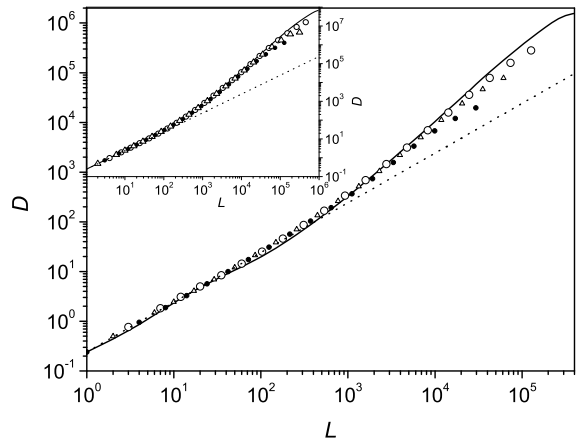}}
\caption{Numerically calculated variance $D(L)$ for the
coarse-grained text of the Bible~\cite{bibe} (solid line) and for
the sequences obtained after their regular decimation. Circles,
triangles, and dots correspond to the decimation parameters 2, 4,
and 8, respectively. The dotted line describes the non-correlated
Brownian diffusion, $D_{0}(L)=L\bar{a}(1-\bar{a})$. The similar
curves obtained for the sequence constructed by using the
long-range part of Bible's memory function only are shown in the
inset.} \label{Fig7}
\end{centering}
\end{figure}

The power-law decrease (without characteristic scale) of the
memory function at long distances leads to quite an essential
property of \textit{self-similarity} of the coarse-grained texts
with respect to the \textit{decimation} procedure discussed in
Ref.~\cite{uya}. This procedure implies the deterministic or
random removal of some part of symbols from a sequence and is
characterized by the decimation parameter $\lambda < 1$ which
represents the fraction of symbols kept in the chain. For example,
under the random decimation each symbol is eliminated with
probability $1-\lambda$. It can be shown that both of these
procedures, deterministic and stochastic, are equivalent for a
Markov chain. The sequence is self-similar if its variance $D(L)$
does not change after the decimation up to a definite value of $L$
(which is dependent on the memory length of the original sequence
and the decimation parameter). The model of the additive binary
many-step Markov chain with the step-like MF (which was discussed
in Ref.~\cite{uya}) offers the exact property of self-similarity
at the length shorter than the memory length $N$. The
coarse-grained literary texts have the self-similarity property as
well. It is indicated in Fig.~\ref{Fig7} where three $D(L)$ curves
correspond to different values of the parameters of the regular
decimation. Note that the decimation procedure leads to a decrease
in the effective memory length. As a result, the variance curves
coincide up to the effective memory depth, which is proportional
to the decimation parameter. A similar phenomenon occurs in the
case of random decimation as well.
\protect

A question arises: what particular property of the memory function
is crucial for the self-similarity of the coarse-grained literary
texts. It is natural to assume that the persistent long-range
scale-free portion of the memory function affords this property
because the self-similarity is specifically manifest at long
distances. To verify this supposition we carry out the decimation
procedure with different $\lambda$ for the Markov chain
constructed by using the long-range part of the Bible memory
function only and then plot the correspondent $D(L)$ dependence.
The curves are shown in the inset in Fig.~\ref{Fig7}. It is seen
that the property of self-similarity for this sequence appears to
be much more pronounced than for the original coarse-grained text
of the Bible. Moreover, the antipersistent part of the MF
disappears very fast after the decimation procedure. This is
clearly observed as a disappearance of the antipersistent
sub-linear portion of the $D(L)$ curves in Fig.~\ref{Fig7} where
after decimation the solid line transforms into the wholly
persistent super-linear curve, which goes above the curve
$D_{0}=L\bar{a}(1-\bar{a})$. The conclusion about the invariance
of the statistical properties of studied sequence with respect to
the decimation procedure is an additional argument in favor of
coarse-graining efficiency. The decimation can be considered as
additional coarse-graining of the initial random sequence.

\section{Conclusion}

Thus, we have demonstrated that the description of the literary
works is suitable in terms of the Markov chains with complex
memory functions. Actually, the memory function appears to be a
convenient informative "visiting card" of any symbolic stochastic
process. We have studied the coarse-grained literary texts and
shown the complexity of their organization in contrast to a
previously discussed simple power-law decrease of correlations. We
have proved that the competition between the two kinds of
correlations govern the statistical properties of the
coarse-grained texts. The antipersistent correlations exist at
short distances, $L \alt 300$, in the region of grammatical rules
efficiency. Another kind of correlations, persistent one, plays
the main role at long distances, $L \agt 300$. It can be related
to the general idea of a literary work. Therefore, the first kind
of correlations may be referred to as the grammatical one, whereas
the second kind may be named as semantic correlations. However,
the nature of the correlations should be clarified by linguists.

If our supposition about the nature of both kinds of correlations
in the literary texts is correct, several important questions will
be of great interest, e.g.:
\begin{itemize}
  \item Does the lack of the antipersistent portion in the memory
  function (and in the $D(L)$ dependence~\cite{uyakm}) in the DNA
  texts mean that the ''grammatical rules'' are absent in the ''DNA
  language''?
  \item If we consider the variance $D(L)$ as a measure of information
redundancy, can we explain the equality $D(L)_{\mid DNA}\simeq
10\cdot D(L)_{\mid Text}$ resulting from the comparison between
literary and DNA texts at $L\sim 3\times 10^5$~\cite{uyakm} in the
following way: the Nature is more careful about the conservation
of the information stored in the DNA sequences than the Writer in
his literary works?
\end{itemize}

We have examined the simplest examples of random sequences, the
dichotomic one. However, our preliminary consideration shows that
the presented concept of additive Markov chains can by generalized
to a larger class of random Markov processes with the finite or
infinite number of states in the discrete or continuous ''time''.
The suggested approach can be used for the analysis of other
correlated systems in different fields of science.


\begin{references}
\bibitem{border} At length $L$ of the order of the full text
length $\mathcal{M} \sim 10^{6}$, we observe a violation of the
monotonous growth of the variance $D(L)$. It is a manifestation of
the border effect. Obviously, $k(L) \rightarrow \overline{k}$ and
the variance $D(L) \rightarrow 0$ (see definition Eq.~(\ref{3}))
at $L \rightarrow \mathcal{M}$.
\bibitem{dif}Note that the antipersistent regions in
Figs.~\ref{Fig1} and \ref{Fig2} differ from each other; this
distinction is connected to the difference in the coarse-graining
procedures used in these two cases.
\end{references}
\end{document}